\documentclass[smallextended]{svjour3}       
\smartqed  
\usepackage{graphicx}
 \journalname{Transport in Porous Media}
\begin{document}

\title{A new set of equations describing immiscible two-phase flow 
in isotropic porous media}

\titlerunning{New equations}     

\author{Alex Hansen \and Santanu Sinha \and Dick Bedeaux \and
Signe Kjelstrup \and Isha Savani \and Morten Vassvik}

\institute{Alex Hansen \and Santanu Sinha \and Isha Savani 
\and Morten Vassvik \at
              Department of Physics\\
              Norwegian University of Science and Technology\\
              NO-7491 Trondheim\\
              Norway\\
              Tel.: +47-73593649\\
              \email{Alex.Hansen@ntnu.no}\\
              \emph{Present address of Santanu Sinha:} CSRC\\
              10 West Dongbeiwang Road\\
              Haidan District, Beijing 100193\\
              China
           \and
              Dick Bedeaux \and Signe Kjelstrup \at
              Department of Chemistry\\
              Norwegian University of Science and Technology\\
              NO-7491 Trondheim\\
              Norway\\
              Tel.: +47-73594179\\
              \email{Signe.Kjelstrup@ntnu.no}\\ 
}

\date{Received: date / Accepted: date}

\maketitle

\begin{abstract}
Based on non-equilibrium thermodynamics we derive a set of general equations 
relating the partial volumetric flow rates to each other and to the total
volumetric flow rate in immiscible two-phase flow in porous media. These 
equations together with the conservation of saturation reduces the 
immiscible two-phase flow problem to a single-phase flow problem of a
complex fluid.  We discuss the new equation in terms of the relative 
permeability equations.  We test the equations on model systems, both 
analytically and numerically.

\end{abstract}

\section{Introduction}
\label{intro}

The simultaneous flow of immiscible fluids through porous media has 
been studied for a long time \cite{b72}. It is a problem that lies at the 
heart of many important geophysical and industrial processes.  Often, the 
length scales in the problem span numerous decades; from the pores 
measured in micrometers to researvoir scales measured in kilometers.  
At the largest scales, the porous medium is treated as a continuum 
governed by effective equations that encode the physics at the pore scale.  

The problem of tying the pore scale physics together with the effective 
description at large scale is the upscaling problem.  In 1936, Wycoff and 
Botset proposed a generalization of the Darcy equation to immiscible 
two-phase flow \cite{wb36}.  It is instructive to reread Wycoff 
and Botset's article.   This is where the concept of relative permeability 
is introduced.  The paper is eighty years old and yet it is still remarkably 
modern. Capillary pressure was first considered by Richards as early as 1931 
\cite{r31}.  In 1940, Leverett combined capillary pressure with the concept 
of relative permeability, and the framework describing essentially all later 
analysis of immiscible multiphase flow in porous media was in place \cite{l40}. 

The introduction of the concepts of relative permeability and capillary 
pressure as solution to the upscaling problem, dominates still today
but other theories exist \cite{lsd81,hg90,hg93a,hg93b,h98,hb00,h06a,%
h06b,h06c,hd10,nbh11,dhh12,habgo15,h15,gsd16}.  These theories are, as is 
relative permeability, based on a number of detailed assumptions concerning 
the porous medium and concerning the physics involved.     

It is the aim of this paper to present a new theory for flow in porous
media that is solely based on a thermodynamic balance between work done
on the flowing fluids and the disspation in them. In the same way as
Buckley and Leverett's analysis based on the conservation of the mass
of the fluids in the porous medium led to their Buckley-Leverett equation 
\cite{bl42}, the balancing of work per time against dissipation
leads to the new equations  we present in the following.

The theory we present rests on non-equilibrium thermodynamics
\cite{kp98,kb08,kbjg10} which combines conservation laws with the 
laws of thermodynamics.  The structure of the theory is reminiscent of 
the structure of thermodynamics itself: we have a number of variables 
that are related through general thermodynamic principles leaving an 
equation of state to account for the detailed physics of the problem.

We only consider here {\it isotropic\/} porous media where the local flow 
always points in the opposite direction of the pressure gradient. It will 
be defered to later to consider more general systems.  We also consider for 
now the pressure  gradient as the only driving force in the system.  

In section \ref{system} we describe the porous medium system we consider.  
We review the key concepts that will be used in the subsequent discussion.
In particular, we discuss the relation between average seepage velocity
and the seepage velocities of each of the two fluids.  We then ask 
the central question: given the average seepage velocity 
and the saturation, is this enough to determine the seepage velocities of 
each of the two fluids?  The answer seems to be ``no," and we demonstrate why.
However, in section \ref{diss} we use non-equilibrium thermodynamics to 
balance dissipation in the two-fluid system against the power supplied to 
it by the pressure difference across it.  We demonstrate that the splitting 
of the dissipation into that of each fluid is unique. In section \ref{scaling} 
we use a scaling argument resting on the Euler theorem for homogeneous 
functions to set up the framework allowing us to derive a set of equations 
relating the average seepage velocity, the saturation and the seepage 
velocities of each of the two fluids.  Section \ref{saturation} contains 
the derivation of these equations.  In section \ref{non-eq} we summarize 
briefly the theory and point out the role played by the constitutive equation.
We derive in section \ref{fractional} a fractional flow 
equation, which essentially is a rewriting of one of the two central
equations in section \ref{saturation}.  Section \ref{solving} contains 
solutions to the two equations of section \ref{saturation}.  We obtain 
integral expressions for the two fluid seepage velocities which are closed 
by combining them with the constitutive equation.  In section \ref{2examples}, 
we demonstrate the use of the formulas derived in the previous 
section to relate a given average seepage velocity as function of saturation 
to the seepage velocities of each of the two fluids.  In section 
\ref{numerical} we analyze data obtained through a numerical pore scale 
network model \cite{amhb98}. Section \ref{relperm} uses the relative
permeability equations as constitutive equations demonstrating that
the equations derived in section \ref{saturation} lead to equations
between the relative permeabilities and between the capillary pressure and 
the relative permeabilities.  This leads to problems for the relative 
permeability formulation as the equations predict that at least one of
the relative permeabilities must depend on the ratio between the viscosities
of the two fluids. We conclude that relative permeability theory is
not thermodynamically consistent.  In section \ref{conclusion} we write 
down the full set of equations to describe immiscible two-phase flow in 
porous media, equations (\ref{conc-1}) to (\ref{conc-7}) based on our theory.  
It reduces the problem from a two-phase fluid problem to a one-phase fluid 
problem with the saturation as a conserved, extra variable. 

\section{Defining the system}
\label{system}

The aim of this paper is to derive a set of equations on the continuum
level where differentials make sense from non-equilibrium thermodynamics.  
We define a {\it representative elementary volume\/} --- REV --- as an 
isotropic block of porous material filled with two immiscible fluids with 
no internal structure:  It is fully described by a small set of parameters 
which we will now proceed to define.  We will then imagine the volume of 
the REV shrinking to zero so that the equations we derive become pointwise 
equations.  These equations 
may then be supplemented by conservation laws to produce a full set of flow 
equations in the same way as is done with the relative permeability
equations. 

\begin{figure*}
\includegraphics[width=1.00\textwidth,clip]{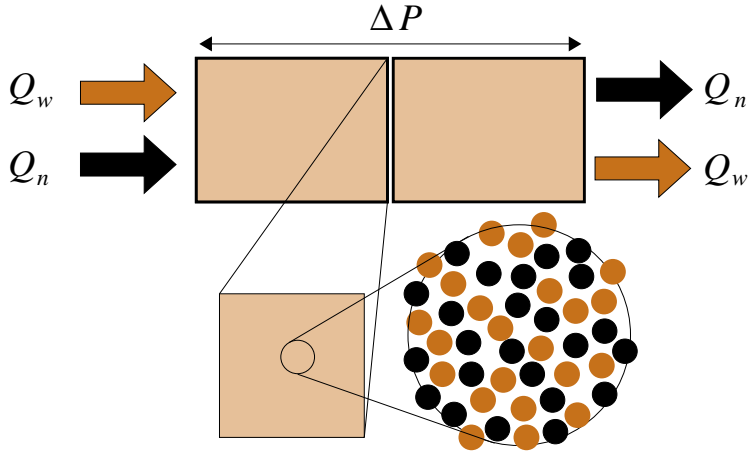}
\caption{In the upper part of the figure, we see the REV from the
side.  A pressure difference $\Delta P$ is applied across it leading to
a flow $Q=Q_w+Q_n$.  An imaginary cut is made through the REV in
the direction orthogonal to the flow.  In the lower left corner, the surface
of the imaginary cut is illustrated.  A magnification of the surface of the
cut is shown in the lower right corner. The pore structure is illustrated as
brown and black circles. The pores that are brown, are filled with wetting
fluid and the pores that are black, are filled with non-wetting fluid. The 
wetting fluid-filled pores form in total an area $A_w$ and the non-wetting
fluid-filled pores form in total an area $A_n$.  The total pore area of
the imaginary cut in the lower left corner is $A_p=A_w+A_n$.}
\label{fig1}   
\end{figure*}

We show in Fig.\ \ref{fig1} the REV. It is a block of homogeneous
porous material of length $L$ and area $A$.  We seal off the surfaces that 
are parallel to the $L$ direction.  The two remaining surfaces, each with 
area $A$, are kept open and act as inlet and outlet for the fluids that are 
injected and extracted from the REV.  The porosity is $\phi=V_p/(AL)$ where 
$V_p$ is the pore volume.\footnote{We define here that $V_p$ is the 
{\it effective\/} pore volume that exclude any irreducible 
wetting fluid or residual non-wetting fluid.} Due to the homogeneity 
of the porous medium, any cut orthogonal to the axis along the $L$ 
direction (named the $x$ axis for later) will reveal a {\it pore area\/} that 
fluctuates around the value $A_p=V_p/L=\phi A$. The homogeneity 
assumption consists in the fluctuations being so small that they can 
be ignored.    

There is a pressure drop $\Delta P$ across the REV as shown in Fig.\ \ref{fig1}.
This leads to a time averaged volumetric flow rate $Q$.  We assume that
the REV is isotropic so that $Q$ --- and other volumetric flow rates --- are 
in the direction of the negative pressure gradient. The volumetric flow rate
consists of two components, $Q_w$ and $Q_n$, which are the volumetric flow
rates of the more wetting ($w$ for ``wetting") and the less wetting ($n$ for 
``non-wetting") fluids with respect to the porous medium.  We have
\begin{equation}
\label{eqn1}
Q=Q_w+Q_n\;.
\end{equation}  

In the porous medium, there is a volume $V_w$ of {\it incompressible\/}
wetting fluid (excluding the irreducible contents of wetting fluid) and a 
volume $V_n$ of {\it incompressible\/}
non-wetting fluid so that $V_p=V_w+V_n$.  We define the wetting and 
non-wetting saturations $S_w=V_w/V_p$ and $S_n=V_n/V_p$. We have that
\begin{equation}
\label{eqn2}
S_w+S_n=1\;.
\end{equation}

We define the wetting and non-wetting pore areas $A_w$ and $A_n$ as the 
parts of the pore area $A_p$ which is filled with the wetting or the 
non-wetting liquids respectively.  As the porous medium is homogeneous, we 
will find the same averages $A_w$ and $A_n$ on any cut 
through the cylindrical porous medium orthogonal to the axis.  This is
illustrated in Fig.\ \ref{fig1}. We have that 
$A_w/A_p=(A_w L)/(A_p L)=V_w/V_p= S_w$ so that
\begin{equation}
\label{eqn3}
A_w=S_w A_p\;.
\end{equation}
Likewise, 
\begin{equation}
\label{eqn4}
A_n=S_nA_p\;.
\end{equation}
Hence, we have
\begin{equation}
\label{eqn5}
A_p=A_w+A_n\;.
\end{equation}

We define the seepage velocities for the two immiscible fluids, $V_w$ and 
$V_n$ as
\begin{equation}
\label{eqn5.1}
v_w=\frac{Q_w}{A_w}\;,
\end{equation}
and
\begin{equation}
\label{eqn5.2}
v_n=\frac{Q_n}{A_n}\;.
\end{equation}
Hence, equation (\ref{eqn1}) may be written    
\begin{equation}
\label{eqn14.9}
Q=A_wv_w +A_nv_n\;.
\end{equation}
We finally define an average seepage velocity associated with the total 
flow rate $Q$ as
\begin{equation}
\label{eqn5.3}
v=\frac{Q}{A_p}\;.
\end{equation}
By using equations (\ref{eqn3}) to (\ref{eqn5}) and 
(\ref{eqn5.1}), (\ref{eqn5.2}) and (\ref{eqn5.3}) we transform 
(\ref{eqn14.9}) into
\begin{equation}
\label{eqn5.4}
v=S_w v_w+S_nv_n\;.
\end{equation}

It is important to note that we assume the REV to be small
enough so that the saturations and seepage velocities may be taken as constant 
throughout the sample.  There is no internal structure in the REV.

\subsection{Non-Uniqueness of equation (\ref{eqn5.4})}
\label{uni}

We pose here the central question that will be answered positively in the
next two sections: {\it from knowledge of $v$ and $S_w$ in equation 
(\ref{eqn5.4}), is it possible to determine $v_w$ and $v_n$, defined in 
equation (\ref{eqn5.1}) and (\ref{eqn5.2})?\/}  Seemingly, the answer is 
``no."  Given a constant reference velocity $v_0$, we may define the new 
velocities
\begin{equation}
\label{eqn5.5}
\tilde{v}_w=v_w+v_0S_n\;,
\end{equation}
and
\begin{equation}
\label{eqn5.6}
\tilde{v}_n=v_n-v_0S_w\;,
\end{equation}
and we have
\begin{equation}
\label{eqn5.7}
A_w\tilde{v}_w+A_n\tilde{v}_n=A_wv_v+A_nv_n=v\;.
\end{equation}
Hence, there is seemingly no way to determine which pair of seepage
velocities corresponds to equation (\ref{eqn5.1}) and (\ref{eqn5.2}) without
knowing $Q_w$ or $Q_n$.  This turns out not to be correct.  By balancing
the dissipation in the REV against the power delivered to the 
REV by the action of the pressure difference $\Delta P$, we arrive at
unique split of the average seepage velocity $v$ into $v_w$ and $v_n$ based 
on the basic laws of thermodynamics.   This is the subject of the next
two sections.

\section{Balancing dissipation and power}
\label{diss}

The following discussion is based on non-equilibrium thermodynamics 
\cite{kbjg10}. 

Let $s=S/V_p$ be the entropy density, $u=U/V_p$ the internal energy density 
and $T$ the temperature of the fluids in the REV. The fluids have
chemical potentials $\mu_w$ and $\mu_n$.  Their molar concentrations are
$c_w=N_w/V_p$ and $c_n=N_n/V_p$ where $N_w$ and $N_n$ are their respective
molar numbers in the REV.  The Gibbs relation between these 
variables is
\begin{equation}
\label{eqn14-1}
Tds=du-\mu_wdc_w-\mu_ndc_n\;,
\end{equation} 
This leads to a rate of change of entropy density with time given by
\begin{equation}
\label{eqn14-2}
\frac{\partial s}{\partial t} = \frac{1}{T}\ \frac{\partial u}{\partial t}
-\frac{\mu_w}{T}\ \frac{\partial c_w}{\partial t}
-\frac{\mu_n}{T}\ \frac{\partial c_n}{\partial t}\;,
\end{equation}
where $t$ is the time variable.  We now introduce conservation equations
for entropy, energy and mass of each fluid,
\begin{equation}
\label{eqn14-3}
\frac{\partial u}{\partial t}+\frac{\partial J_u}{\partial x}=0\;,
\end{equation}
\begin{equation}
\label{eqn14-4}
\frac{\partial c_w}{\partial t}+\frac{\partial J_w}{\partial x}=0\;,
\end{equation}
and
\begin{equation}
\label{eqn14-5}
\frac{\partial c_n}{\partial t}+\frac{\partial J_n}{\partial x}=0\;,
\end{equation}
where the $x$ axis is oriented along the positive flow direction of the REV,
see Fig.\ \ref{fig1}.
Here $J_u$ is the current density of energy. The other two are the 
current densities of the wetting fluid and non-wetting fluid molar numbers 
given by 
\begin{equation}
\label{eqn14-11}
J_w=c_w \overline{v}_w\;,
\end{equation}
and
\begin{equation}
\label{eqn14-12}
J_n=c_w \overline{v}_n\;,
\end{equation}
where $\overline{v}_w$ and $\overline{v}_n$ are the velocities associated
with these currents.  By combining the conservation laws (\ref{eqn14-3}) to
(\ref{eqn14-5}) with (\ref{eqn14-2}) we find
\begin{equation}
\label{eqn14-6}
\frac{\partial s}{\partial t}=
-\frac{\partial}{\partial x}\ \frac{1}{T}\
\left[J_u-J_w\mu_w-J_n\mu_n\right]+J_u\ \frac{\partial}{\partial x}\ 
\frac{1}{T}\ 
-J_w\ \frac{\partial}{\partial x}\ \frac{\mu_w}{T}
-J_n\ \frac{\partial}{\partial x}\ \frac{\mu_n}{T}\nonumber\\
\end{equation}
If we now compare this expression to the entropy balance equation
\begin{equation}
\label{eqn14-7}
\frac{\partial s}{\partial t}+\frac{\partial J_s}{\partial x}=\sigma\;,
\end{equation}
where $J_s$ is the entropy current density and $\sigma$ is the entropy
production density, we find
\begin{equation}
\label{eqn14-8}
J_s=\frac{1}{T}\left[J_u-J_w\mu_w-J_n\mu_n\right]\;,
\end{equation}
and
\begin{equation}
\label{eqn14-9}
\sigma=J_u\ \frac{\partial}{\partial x}\ \frac{1}{T}
-J_w\ \frac{\partial}{\partial x}\ \frac{\mu_w}{T}
-J_n\ \frac{\partial}{\partial x}\ \frac{\mu_n}{T}\;.
\end{equation}
The system is in a steady state so that 
$\partial s/\partial t=0$.  We also assume that the temperature $T$ is 
constant throughout the REV.  The steady state 
assumption also leads to 
$\partial J_u/\partial x=\partial J_w/\partial x=\partial J_n/\partial x=0$
through equations (\ref{eqn14-3}) to (\ref{eqn14-5}).  We integrate
the expression for the entropy production (\ref{eqn14-9}) over the pore volume 
of the REV, giving the dissipation $D$ in the system,
\begin{equation}
\label{eqn14-10}
D=A_p\ \int_0^L dx\ T\sigma=A_p\left[J_w\Delta\mu_w+
J_n\Delta\mu_n\right]=
A_p\left[c_w\Delta\mu_w \overline{v}_w+c_n\Delta\mu_n \overline{v}_n\right]\;,
\end{equation}
where $\Delta\mu_w=\mu_w(0)-\mu_n(L)$ and $\Delta\mu_n=\mu_w(0)-\mu_n(L)$ 
are the chemical potential differences between the entrance and the exit
of the REV for the fluids. The dissipation $D$ must equal the work done on 
the system by the pressure difference $\Delta P$ per time unit, $Q\Delta P$.  
The power supplied to the wetting and the non-wetting fluids by the pressure 
gradient is $\overline{Q}_w\Delta P$ and $\overline{Q}_n\Delta P$ respectively,
where $\overline{Q}_w=V_w\overline{v}_w$ and 
$\overline{Q}_n=V_n\overline{v}_n$. The power supplied to each fluid must
balance the dissipation in each fluid if the system is to be in a steady
state.  Hence,
\begin{equation}
\label{eqn14-13}
V_w\overline{v}_w\Delta P=V_p c_w\Delta\mu_w\overline{v}_w\;,
\end{equation}
and 
\begin{equation}
\label{eqn14-14}
V_n\overline{v}_n\Delta P=V_p c_n\Delta\mu_n\overline{v}_n\;.
\end{equation}
Combining these two equations with (\ref{eqn14-10})
gives
\begin{equation}
\label{eqn14-15}
Q\Delta P=A_p v\Delta P=
D=A_p\left[S_w \overline{v}_w+S_n\overline{v}_n\right]\Delta P\;,
\end{equation}
where we have used that $V_w/V_p=S_w$ and $V_n/V_p=S_n$. Hence, we
have 
\begin{equation}
\label{eqn15-1}
v=\left[S_w \overline{v}_w+S_n\overline{v}_n\right]\;,
\end{equation}
in analogy to equation (\ref{eqn5.4}).  What has been accomplished here is 
(1) a unique split of the average seepage velocity $v$ into two velocities
$\overline{v}_w$ and $\overline{v}_n$ associated with the dissipation in 
each fluid and (2) we have shown that $Q$ is a thermodynamic variable. 

As we shall see in section \ref{numerical}, in small systems $\overline{v}_w$ 
and $\overline{v}_n$ defined in equations (\ref{eqn14-11}) and 
(\ref{eqn14-12}), are not identical to the seepage velocities defined in
equations (\ref{eqn5.1}) and (\ref{eqn5.2}) since they are different averages.
However, in the thermodynamic limit which is the continuum limit, they will 
be equal due to the system being self averaging, and we have
\begin{equation}
\label{eqn15-2}
\overline{v}_w=v_w=\frac{Q_w}{A_w}\;,
\end{equation}
and 
\begin{equation}
\label{eqn15-3}
\overline{v}_n=v_n=\frac{Q_n}{A_n}\;.
\end{equation}

\section{Scaling assumption}
\label{scaling}

\begin{figure*}
\includegraphics[width=1.00\textwidth,clip]{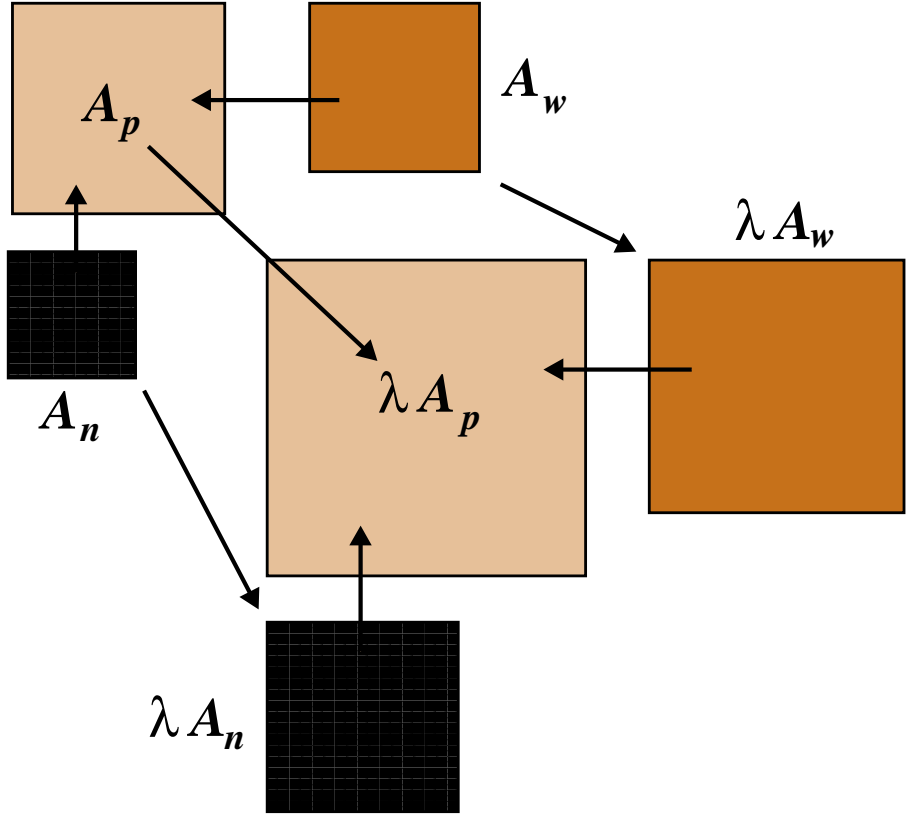}
\caption{In the upper left corner, we show an imaginary cut through the
REV in the direction orthogonal to the flow as shown in the lower
left corner of Fig.\ \ref{fig1}.
The pore area is $A_p$.  This is decomposed in the area covered by the
wetting fluid $A_w$ and the non-wetting fluid $A_n$.  These areas 
are shown below ($A_w$) and to the right ($A_n$) of the cut in the 
upper left corner.  In the center, we show the {\it rescaled\/}
imaginary cut.  That is, we imagine the original REV replaced by another
rescaled one.  The rescaling has been done with a factor $\lambda$.  Hence,
we have $A_p\to\lambda A_p$, and consequently, $A_w\to\lambda A_w$ and
$A_n\to\lambda A_n$ as show below and to the right of the rescaled cut.
The length of the REV, $L$ is {\it not\/} rescaled.}
is,
\label{fig2}   
\end{figure*}

The volumetric flow rate $Q$, which we now recognize as a thermodynamic
function, is a {\it homogeneous function of order one\/} 
of the two areas $A_w$ and $A_n$ defined in equations (\ref{eqn3}) and 
(\ref{eqn4}).  Hence, if $\lambda$ is a scale factor, we may scale the 
two areas $A_w\rightarrow \lambda A_w$ and $A_n\rightarrow \lambda A_n$. We
illustrate this in Fig.\ \ref{fig2}.  This leads to the scaling relation
\begin{equation}
\label{eqn10}
Q(\lambda A_w,\lambda A_n)=\lambda Q(A_w,A_n)\;.
\end{equation}
This scaling property is essentially self evident. It may be implemented in
pratice by using different porous media samples with different areas $A$. 

We take the derivative with respect to $\lambda$ on both sides of 
(\ref{eqn10}) and set $\lambda=1$. This gives
\begin{equation}
\label{eqn11}
Q(A_w,A_n)=A_w \left(\frac{\partial Q}{\partial A_w}\right)_{A_n}+
A_n \left(\frac{\partial Q}{\partial A_n}\right)_{A_w}\;,
\end{equation}
where it is understood that we keep the pressure drop $\Delta P$ 
constant during the partial differentiations.  Equation
(\ref{eqn11}) is essentially the Euler theorem for homogeneous functions of
order one.  By dividing this equation by $A_p$, we have
\begin{equation}
\label{eqn10-1}
v=S_w \left(\frac{\partial Q}{\partial A_w}\right)_{A_n}+
S_n \left(\frac{\partial Q}{\partial A_n}\right)_{A_w}\;.
\end{equation}
By comparing this equation to (\ref{eqn15-1}) using (\ref{eqn15-2})
and (\ref{eqn15-3}), we find
\begin{equation}
\label{eqn10-2}
v_w=\frac{Q_w}{A_w}=\left(\frac{\partial Q}{\partial A_w}\right)_{A_n}\;,
\end{equation}
and
\begin{equation}
\label{eqn10-3}
v_n=\frac{Q_n}{A_n}=\left(\frac{\partial Q}{\partial A_n}\right)_{A_w}\;.
\end{equation}

\section{Dependence on saturation: new equations}
\label{saturation}

Let us return to the scaling equation (\ref{eqn10}) which we combine with the
expression for the total flow in terms of the seepage velocities $v_w$ and
$v_n$, equation (\ref{eqn14.9}).
In order for $Q$ in this equation to obey the scaling relation (\ref{eqn10}),
neither $v_w$ nor $v_n$ can be extensive in the variables $A_w$ and $A_n$, 
i.e., 
\begin{eqnarray}
\label{eqn14a}
& v_w(\lambda A_w,\lambda A_n) = v_w(A_w,A_n)\;,\nonumber\\
& v_n(\lambda A_w,\lambda A_n) = v_n(A_w,A_n)\;,\nonumber\\
& v  (\lambda A_w,\lambda A_n) = v  (A_w,A_n)\;.\nonumber\\
\end{eqnarray}
They are homogeneous functions of $A_w$ and $A_n$ of order zero.
Hence, they must depend on the areas $A_w$ and $A_n$ through their
ratio $A_w/A_n=S_w/S_n=S_w/(1-S_w)$, where we have used equations 
(\ref{eqn3}) --- (\ref{eqn5}). Hence,
\begin{eqnarray}
\label{eqn14b}
& v_w=v_w(S_w)\;,\nonumber\\
& v_n=v_n(S_w)\;,\nonumber\\
& v  =v  (S_w)\;.\nonumber\\
\end{eqnarray}

The scaling relation (\ref{eqn10}) implies changing the wetting and 
non-wetting pore areas $A_w$ and $A_n$ by changing $A_p$, but without 
changing the saturations $S_w=A_w/A_p$ and $S_n=A_n/A_p$.  We now
change the saturation while keeping the total pore area $A_p=A_w+A_n$ constant.
Mathematically, this is accomplished by changing our variables from $(A_w,A_n)$
to $(S_w,A_p)$ where $A_w=S_w A_p$
and $A_n=(1-S_w)A_p$. 

We calculate 
\begin{equation}
\label{eqn18.0}
\left(\frac{\partial Q}{\partial S_w}\right)_{A_p} =
\left(\frac{\partial A_w}{\partial S_w}\right)_{A_p} 
\left(\frac{\partial Q}{\partial A_w}\right)_{A_n} +
\left(\frac{\partial A_n}{\partial S_w}\right)_{A_p} 
\left(\frac{\partial Q}{\partial A_n}\right)_{A_w}
= A_p\left[v_w-v_n\right]\;,
\end{equation}
where we have used (\ref{eqn10-2}) and (\ref{eqn10-3}). We divide by the
area $A_p$ which is kept constant and find
\begin{equation}
\label{eqn18}
\frac{dv}{dS_w}=v_w-v_n\;.
\end{equation}
This equation is a direct consequence of scaling assumption (\ref{eqn10}).
This equation is one of the two fundamental equations that 
constistute the main result presented in this paper.

The second equation we find by taking the derivative of equation (\ref{eqn5.4})
with respect to $S_w$,
\begin{equation}
\label{eqn19}
\frac{dv}{dS_w}=\frac{d}{dS_w}\left[S_w v_w+(1-S_w)v_n\right]
=v_w-v_n+S_w \frac{dv_w}{dS_w}+(1-S_w)\frac{dv_n}{dS_w}\;.
\end{equation}
Combining this equation with equation (\ref{eqn18}), we have the equation
\begin{equation}
\label{eqn20}
S_w \frac{dv_w}{dS_w}+(1-S_w)\frac{dv_n}{dS_w}=0\;,
\end{equation}
which is our second main result. Also this equation is a consequence of the 
scaling assumption (\ref{eqn10}) alone and does not entail any further 
assumpions about the properties of the flow problem.

Equation (\ref{eqn20}) is in fact the {\it Gibbs-Duhem equation,\/} in 
disguise, see e.g.\ Reference \cite{kp98}.  

Let us lastly in this section remark that the Darcy or superficial velocities
$V$, $V_w$ and $V_n$ are related to the seepage velocities by $V=v\phi$,
$V_w=v_w\phi$ and $V_n=v_n\phi$.  Hence, equations (\ref{eqn18}) and
(\ref{eqn20}) will not change if expressed in terms of these
velocities instead of the seepage velocities.

\section{Non-equilbrium thermodynamic description}
\label{non-eq}

We note that the three equations (\ref{eqn5.4}), (\ref{eqn18}) and
(\ref{eqn20}) are related in such a way that given any two of them, the 
third follows.

We have accomplished to construct a non-equilibrium thermodynamic 
theory for immiscible two-phase flow in porous media \cite{kb08,kbjg10}.  There
are three velocities, $v$, $v_w$ and $v_n$.  These are the responses to
the pressure difference $\Delta P$.  Lastly, there is the saturation.
The three velocities are bound together by the two equations (\ref{eqn18}) 
and (\ref{eqn20}).  Hence, three variables remain: say $v$, $S_w$ and 
$dP/dx=P'=\Delta P/L$.  They will be related through a 
{\it constitutive equation\/}
\begin{equation}
\label{eqn20.1}
v=v(S_w,P');.
\end{equation}
It is through this equation that the detailed physics enters the description.
We note that this equation plays the same role as an {\it equation of state\/}
in thermodynamics.  The equations we have derived refer only to velocities,  
saturation and pressure gradient.  Hence, the non-zero volume of the REV 
has dropped out and the two equations may be seen as applying pointwise in 
a continuous porous medium.   If the three equations (\ref{eqn18}), 
(\ref{eqn20}) and (\ref{eqn20.1}) are supplied with a conservation law for 
the saturation, a complete descpription of porous media in the continuum
limit ensues. We elaborate on this in the conclusion and discussion 
section \ref{conclusion}.

\section{Fractional flow equation}
\label{fractional}

We define the wetting and non-wetting fractional flows as
\begin{equation}
\label{eqn21}
F_w=\frac{Q_w}{Q}=S_w\frac{v_w}{v}\;,
\end{equation}
and
\begin{equation}
\label{eqn22}
F_n=1-F_w=\frac{Q_n}{Q}=S_n\frac{v_n}{v}\;,
\end{equation}
where we have used equations (\ref{eqn5.1}) and (\ref{eqn5.2}).  We now 
combine these expressions for $F_w$ and $F_n$ with equation (\ref{eqn18}) 
to find
\begin{equation}
\label{eqn23}
\frac{dv}{dS_w}=v\left[\frac{F_w}{S_w}-\frac{F_n}{S_n}\right]\;.
\end{equation}
By solving for $F_w$, we find
\begin{equation}
\label{eqn27}
F_w=S_w+S_w(1-S_w)\ \frac{1}{v}\ 
\frac{dv}{dS_w}\;.
\end{equation}

Hence, we have derived a new {\it fractional flow equation\/} relating the 
fractional flow to the total flow rate and the saturation.  Again, no further 
assumptions have been made concerning the flow apart from the scaling relation 
given in (\ref{eqn10}).  All that is needed to determine the fractional flow 
is contained in the average seepage velocity $v=v(S_w)$, i.e., in the
constitutive law (\ref{eqn20.1}).

\section{Solving the equations}
\label{solving}

Equations (\ref{eqn18}) and (\ref{eqn20}) may be integrated to find $v_w$ and 
$v_n$ as functions
of $S_w$. We start by transforming equation (\ref{eqn18}) into
\begin{equation}
\label{eqn35}
S_w\ \frac{d^2v}{dS_w^2}
= S_w\ \frac{dv_w}{dS_w} - S_w\ \frac{dv_n}{dS_w}\;.
\end{equation}
Equation (\ref{eqn20}) may then be subtracted from this equation to give
\begin{equation}
\label{eqn36}
\frac{dv_n}{dS_w} = - S_w\ \frac{d^2v}{dS_w^2}\;.
\end{equation}
By using equation (\ref{eqn20}) once again, we find
\begin{equation}
\label{eqn37}
\frac{dv_w}{dS_w} = S_n\ \frac{d^2v}{dS_w^2}\;.
\end{equation}

We integrate equations (\ref{eqn36}) and (\ref{eqn37}) finding
\begin{equation}
\label{eqn40}
v_w(S_w)=v_w(1)-\int_{S_w}^1\ dS\ (1-S)\ 
\frac{d^2v}{dS^2}\;,
\end{equation}
and
\begin{equation}
\label{eqn41}
v_n(S_w)= v_n(0)-\int_0^{S_w}\ dS\ S\ \frac{d^2v}{dS^2}\;.
\end{equation}

As with the fractional flow equation (\ref{eqn27}), only the average 
seepage velocity $v=v(S_w)$ is needed to determine seepage velocities for each
of the fluids.  This is no surprise as $F_w$ and $v_w$ convey 
the same information about the system.

\section{Two examples}
\label{2examples}

We now give a couple of analytically tractable examples of the use of 
equations (\ref{eqn40}) and (\ref{eqn41}).  

First a trivial example. Suppose that surface tension between the two 
fluids is negligible. The total average seepage velocity (\ref{eqn5.4}) is 
$v(1)=v_w(1)$ for $S_w=1$ and it is $v(0)=v_n(0)$ for $S_w=0$. If the 
wetting fluid has a viscosity $\mu_w$ and the non-wetting fluid a viscosity 
$\mu_n$, we have $v_w(1)/v_n(0)=\mu_n/\mu_w$.  $v(S_w)$ must 
be linear in $S_w$: $v(S_w)=A+BS_w$ and we must have $A=v_n(0)$ and 
$B=v_w(1)-v_n(0)$ so that
\begin{equation}
\label{eqn42}
v(S_w)=v_n(0)+(v_w(1)-v_n(0))S_w=v_n(0)\left[1+\left(\frac{\mu_n}{\mu_w}-1
\right)S_w\right]\;.
\end{equation}
Equations (\ref{eqn40}) and (\ref{eqn41}) give since $d^2v/dS_w^2=0$
\begin{equation}
\label{eqn43}
v_w(S_w)=v_w(1)\;,
\end{equation}
and 
\begin{equation}
\label{eqn44}
v_n(S_w)=v_n(0);.
\end{equation}
as it should be.

\begin{figure*}
\includegraphics[width=0.75\textwidth,clip]{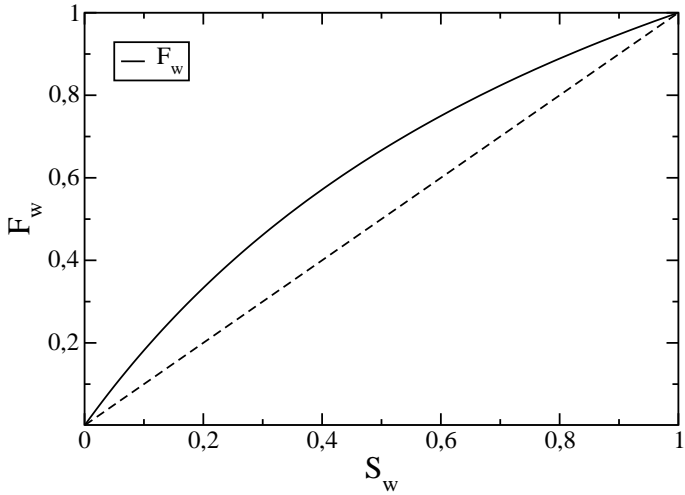}
\caption{Fractional flow $F_w$ as a function of saturation $S_w$ 
for the example given in (\ref{eqn42}) using equation (\ref{eqn27}). 
We have set $\mu_n/\mu_w=2$. The broken line is the diagonal $F_w=S_w$.}
\label{fig3}   
\end{figure*}

We may also test the fractional flow equation (\ref{eqn27}) for this simple
case.  Using $v(S_w)$ from (\ref{eqn42}), we find
\begin{equation}
\label{eqn42.1}
F_w=S_w+\frac{S_w(1-S_w)}{1+(\mu_n/\mu_w-1)S_w}\ 
\left(\frac{\mu_n}{\mu_w}-1\right)\;,
\end{equation}
which is equal to $S_wv_w/v(S_w)$ as it should be.
We illustrate $F_w$ in Fig.\ \ref{fig3}.

We know from equation (\ref{eqn18}) that when $dv/dS_w=0$, $v_w=v_n$.
From equation (\ref{eqn5.4}) this implies furthermore that $v=v_w=v_n$.
As a consquence of (\ref{eqn27}), we also have that $F_w=S_w$
when $dv/dS_w=0$. Let us name this particular saturation $S_{w,m}$. 
When we move away from $S_{w,m}$, we have that
\begin{equation}
\label{eqn42.2}
v(S_w)=v(S_{w,m})+{\cal O}(\delta S_w^2)\;,
\end{equation}
where $\delta S_w=S_w-S_{w,m}$.
Hence, to first order in $\delta S_w$ the two fluids behave as if they were 
miscible and both having the same effective viscosity.  However, the
constitutive equation (\ref{eqn20.1}) describing the mixture may still be more 
complicated than the Darcy equation even if the two fluids each are Newtonian 
\cite{tkrlmtf09,tlkrfm09,rcs11,sh12,sbdkpbtsch16}.

In the example just discussed, there is no minimum in the average
seepage velocity $v$, as it is a constant with respect to $S_W$. 
Hence, as long as $\mu_n \neq \mu_w$, $v_w$ is different
from $v_n$ and --- as seen in Fig.\ \ref{fig3}, $F_w\neq S_w$ for all
values of $S_w$ except the trivial values 0 and 1.

We now move on to our next example.
Let us suppose that the average seepage velocity $v$ may be parametrized
as a fourth order polynomial in $S_w$,
\begin{equation}
\label{eqn47.1}
v(S_w)=\sum_{j=0}^4\ a_j S_w^j\;.
\end{equation}
We find from the integrals (\ref{eqn40}) and (\ref{eqn41})
\begin{equation}
\label{eqn47.2}
v_w(S_w)=(a_0+a_1)+2a_2S_w-(a_2-3a_3)S_w^2-(2a_3-4a_4)S_w^3-3a_4S_w^4\;,
\end{equation}
and
\begin{equation}
\label{eqn47.3}
v_n(S_w)=a_0-a_2S_w^2-2a_3S_w^3-3a_4S_w^4\;.
\end{equation}

\begin{figure*}
\includegraphics[width=0.75\textwidth,clip]{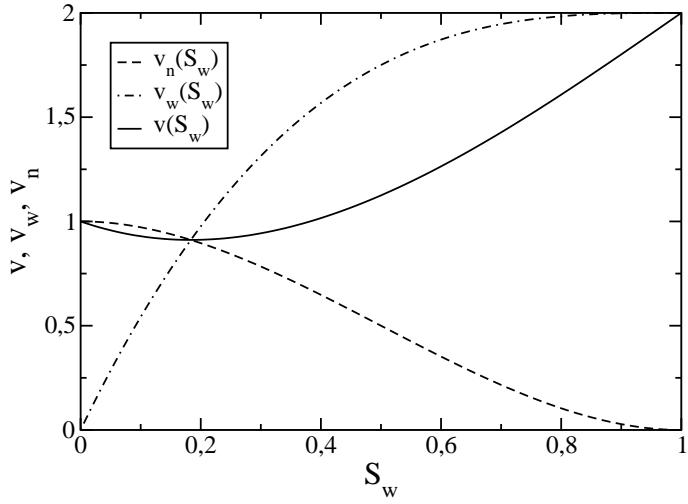}
\caption{Seepage velocities $v_n$ and $v_w$ as calculated from
the average seepage velocity $v$ --- equation (\ref{eqn45}) --- using
equations (\ref{eqn40}) and (\ref{eqn41}). We have set $v_n(0)=1$
and $\mu_n/\mu_w=2$.}
\label{fig4}   
\end{figure*}

A concrete example where $a_4=0$ is
\begin{equation}
\label{eqn45}
v(S_w)=v_n(0)\left[1+\left(\mu_n/\mu_w-1\right)S_w - 
(2-S_w)(1-S_w)S_w\right]\;,
\end{equation}
see Fig.\ \ref{fig4}.
Integrals (\ref{eqn40}) and (\ref{eqn41}) give
\begin{equation}
\label{eqn46}
v_n(S_w)=v_n(0)\left[1-S_w^2(3-2S_w)\right]\;,
\end{equation}
and
\begin{equation}
\label{eqn47}
v_w(S_w)=v_n(0)\left[\mu_n/\mu_w-2(1-S_w)^3\right]\;.
\end{equation}

\begin{figure*}
\includegraphics[width=0.75\textwidth,clip]{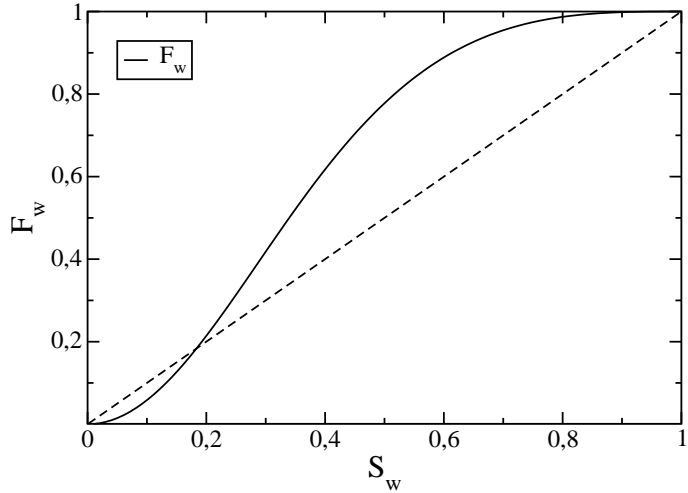}
\caption{Fractional flow $F_w$ as a function of saturation $S_w$ 
for the example given in (\ref{eqn45}) using equation (\ref{eqn27})
with $\mu_n/\mu_w=2$. The broken line is the diagonal $F_w=S_w$.}
\label{fig5}   
\end{figure*}

Fig.\ \ref{fig4} shows $v$, $v_w$ and $v_n$ as function of $S_w$ for 
$\mu_n/\mu_w=2$. We note that the three seepage velocities meet as
a point.  Such a point must exist when $v_w$ is an increasing function and
$v_n$ is a decreasing function of $S_w$ as they must cross somewhere,
and at the point they meet, $v$ must be equal to the two others due to
equation (\ref{eqn5.4}).  However, equation (\ref{eqn18}) dictates that this
point is also the point at which $v$ is minimal. In this case this happens for
$S_{w,m}=1-\sqrt{2/3}$.  We see that this is the case in Fig.\ \ref{fig4}.
We show in Fig.\ \ref{fig5} the fractional flow $F_w$ calculated from
equation (\ref{eqn27}) as a function of $S_w$
for the example given in (\ref{eqn45}).  For the value $S_{w,m}=1-\sqrt{2/3}$,
$F_w=S_w$.  As described above, the two immiscible fluids with different
viscosities act as if they were miscible and have the same viscosity.

\section{Numerical studies}
\label{numerical}

Typically, part of the wetting fluid will be stuck and does not contribute 
to the flow properties apart from lowering the available pore volume.  The 
volume of this bound wetting fluid divided by the pore volume is the 
irreducible wetting saturation $S_{wi}$ and the residual non-wetting
saturation $S_{nrw}$.  It is then convenient to define an effective saturation
\begin{equation}
\label{eqn2.1}
S^*_w=\frac{S_w-S_{wi}}{1-S_{wi}-S_{nrw}}\;.
\end{equation}
We define the corresponding effective non-wetting saturation as 
$S^*_n=1-S^*_w$.  

The network simulator first proposed by Aker et al.\ \cite{amhb98} has been 
refined over the years and is today a versatile model for immiscible 
two-phase flow under steady-state conditions.  The model 
tracks the interfaces between the immiscible fluids by solving the Kirchhoff 
equations with a capillary pressure given by 
\begin{equation}
\label{eq:yl}
p_c = \pm \sum_i^n \frac{2 \gamma}{r} \left(1 - \cos (2 \pi x_i)\right)
\end{equation}
where $x_i$ is the position of the $i$th interface in a link measured in units 
of the length $l$ of the link, $r$ its average radius and $\gamma$ is 
the surface tension between the two fluids.  In the following, we will
use this model to test the fractional flow equation (\ref{eqn27}).

Our parameters are chosen as follows: we have set $l=1$ mm and
$\gamma=0.03$ N/m.  The link radii $r$ were drawn from an approximately 
lognormal distribution with an average of $0.133l$ and a standard deviation 
of $0.044l$. The viscosities of the fluids were equal,
$\mu_w = \mu_n = 0.1$ Pa s. Steady state is obtained by implementing the 
network on a torus.  Hence, the system is closed and the saturation $S_w$ does 
not fluctuate.  We have used a two-dimensional hexagonal lattice consisting of
$100 \times 50$ nodes.  The system is driven by a constant pressure difference 
of 15 kPa. The capillary number was hovering around 0.02.

\begin{figure*}
\includegraphics[width=0.75\textwidth,clip]{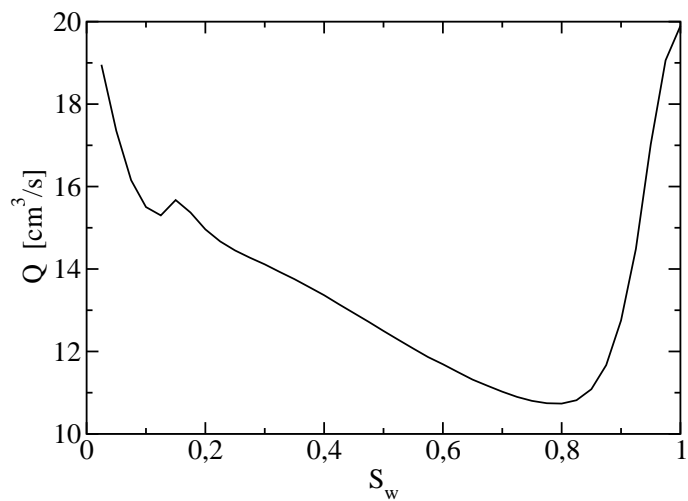}
\caption{Volumetric flow rate $Q$ as a function of $S_w$.}
\label{fig6}   
\end{figure*}

\begin{figure*}
\includegraphics[width=0.75\textwidth,clip]{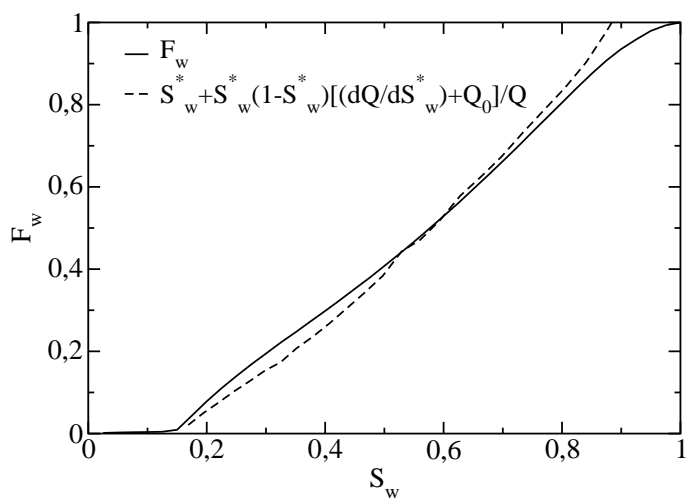}
\caption{Fractional flow rate $F_w$ and equation (\ref{eqn25-3}).}
\label{fig7}   
\end{figure*}

We show in Fig.\ \ref{fig6} the volumetric flow rate $Q$ as a function of $S_w$ 
and in Fig.\ \ref{fig7} the corresponding fractional flow rate $F_w$ as a 
function of $S_w$. Our aim is to compare the fractional flow equation
(\ref{eqn27}) to the calculated fractional flow. 

In the derivation of the fractional flow equation (\ref{eqn27}), the
velocities defined in equations (\ref{eqn15-2}) and (\ref{eqn15-3}),
$\overline{v}_w$ and $\overline{v}_n$ are
used. In a network model, one would measure e.g.\ $\overline{v}_w$ as follows:
let $q_{w,i}$ be the volumetric flow rate of the wetting fluid in the $i$th 
link and $a_i$ is the area of this link. Then, the wetting fluid  velocity 
in this link is $v_{w,i}=q_{w,i}/a_i$. We then average these velocities
over the links. Hence, we have
\begin{equation}
\label{eqn25-1}
\overline{v}_w=\frac{\sum_i v_{w,i}}{N_L N_A} =
\frac{\sum_i \frac{q_{w,i}}{a_i}}{N_L N_A} \;,
\end{equation}
where the sum runs over the all the links in the network.  There are 
$N_A$ links in each layer (row of links in the direction orthogonal to the
flow direction) and $N_L$ layers.

The seepage velocity, $v_w$, defined in (\ref{eqn5.1}) on the other hand,
is calculated in the network as
\begin{equation}
\label{eqn25-2}
v_w= \frac{1}{N_L}\ \sum_k \left(
\frac{\sum_{i\in k} q_{w,i}}{\sum_{j \in k} a_j}\right)\;,
\end{equation}
where the sum index runs over layers in the network in the flow direction. 

In the limit of an infinitely large network the two averages (\ref{eqn25-1}) 
and (\ref{eqn25-2}) will be equal as stated in \ref{diss}.  In a small system, 
such as the one we consider here, they are not, $v_w\neq\overline{v}_w$.  
We measure in Fig.\ \ref{fig7} $v_w$ and not 
$\overline{v}_w$.\footnote{$Q_w$ is measured by keeping track of how much 
of the wetting fluid has crossed a cross section of the system over a time 
step.  $F_w$ is then the ratio of how much non-wetting fluid has passed that 
cross section divided by how much total fluid has passed.}

If we assume that the difference between $v_w$ and $\overline{v}_w$ is of the 
form (\ref{eqn5.5}),  $\overline{v}_w=v_w+v_0S_n$,  
we find the following modified fractional flow equation
\begin{equation}
\label{eqn25-3}
F_w={S^*}_w+{S^*}_w(1-{S^*}_w)\ \frac{1}{Q}\ \left[\frac{dQ}{d{S^*}_w}+Q_0
\right]\;.
\end{equation}
We have defined $Q_0=A_p v_0$. Furthermore, we have used the effective 
saturation ${S^*}_w$ rather than the saturation $S_w$.  We have set 
$S_{wi}=0.15$, $S_{nrw}=0$ and $Q_0=4.5$ cm$^3$/s.  
The derivative $dQ/dS_w$ was calculated by forwards difference with 
$\Delta S_w=0.025$.

The difference between the measured and calculated fractional flow curves
in Fig.\ \ref{fig7} is then due to finite-size effects in the averaging 
process.

\section{Consequences for relative permeability}
\label{relperm}

Theories of immiscible two-phase flow in porous media attempt to provide
a set of equations describing the flow at the continuum scale while anchoring
the key concepts at the pore level.  As mentioned in the introduction, there
are several different theories in existence.  There is the percolation-based
approach \cite{lsd81,gsd16}, there is the approach based on continuum mixture
theory where the interface energies play a major role 
\cite{h98,hb00,h06a,h06b,h06c,hd10,nbh11,dhh12,habgo15}, and there is
the approach where additional variables describing the theormodynamics of
the interfaces are introduced \cite{hg90,hg93a,hg93b,h15}.  The classical,
and by far dominating theory in use today is relative permeability theory
\cite{wb36} even though it is well known to have serious weaknesses.

The two equations (\ref{eqn18}) and (\ref{eqn20}) are based on general
non-equilibrium thermodynamical arguments that only invoke scale symmetry
(\ref{eqn10}) and a balance of dissipation and power, they
must apply to all these theories.  In the following, we analyze relative
permeability theory in light of our equations.

The relative permability approach consists in making explicit assumptions 
about the functional form of $v_w$ and $v_n$ through the generalized Darcy 
equations.  Returning to the REV and the definitions of section \ref{system},
we have
\begin{equation}
\label{eqn7}
v_w=\frac{1}{\phi S_w}\ \frac{Q_w}{A}=-\frac{1}{\phi S_w}\ 
\frac{K}{\mu_w}\ k_{r,w}\ P'_w\;,
\end{equation}
and
\begin{equation}
\label{eqn8}
v_n=\frac{1}{\phi S_n}\ \frac{Q_n}{A}=-\frac{1}{\phi S_n}\ 
\frac{K}{\mu_n}\ k_{r,n}\ P'_n\;.
\end{equation}
Here $K$ is the permeability of the porous medium, and $k_{r,w}$ and 
$k_{r,n}$ are the relative permeabilities of the wetting and non-wetting
fluids.  One distinguishes between the pressure gradient in the wetting fluid
$P'_w=\Delta P_w/L$ and in the non-wetting fluid $P'_n=\Delta P_n/L$.  They
are related through the capillary pressure $P_c$ by
\begin{equation}
\label{eqn9}
P_n-P_w=P_c\;.
\end{equation} 

The assumptions concerning the physical properties of the flow made in the 
relative permeability description are that the relative permeabilities and the 
capillary pressure are functions of the wetting (or equivalently, the 
non-wetting) saturation {\it only,\/} 
$k_{r,w}=k_{r,w}(S_w)$, $k_{r,n}=k_{r,n}(S_w)$ and $P_c=P_c(S_w)$.  

These are strong assumptions, and there is growing evidence that they do 
not hold in general, see 
\cite{tkrlmtf09,tlkrfm09,rcs11,sh12,yts13,bt13,mfmst15}: 
the relative permeabilities and the capillary pressure are {\it not\/} 
functions of the saturation alone. 

Let us now turn to the second fundamental equation (\ref{eqn20}), combining 
it with the relative permeability and capillary pressure equations 
(\ref{eqn7})--(\ref{eqn9}).  By inserting the expressions for $v_w$ and 
$v_n$ in terms of the relative permeabilities, and eliminating $P_w$ by 
equation (\ref{eqn9}), we find 
\begin{eqnarray}
\label{eqn31}
\left[\frac{S_w}{\mu_w}\ \left(\frac{d}{dS_w}\right)
\left(\frac{k_{r,w}}{S_w}\right)+
\frac{1-S_w}{\mu_n}\ \left(\frac{d}{dS_w}\right)
\left(\frac{k_{r,n}}{1-S_w}\right)\right]\
\Delta P_n&
\nonumber\\
-\left[\frac{S_w}{\mu_w}\ \left(\frac{d}{dS_w}\right)
\left(\frac{k_{r,w}}{S_w}\ P_c\right)
\right]&=0\;.\nonumber\\
\end{eqnarray}
This equation must be valid for all values of $\Delta P_n$. Hence, it splits 
into two equations,
\begin{equation}
\label{eqn32}
\frac{S_w}{\mu_w}\
\left(\frac{d}{dS_w}\right) \left(\frac{k_{r,w}}{S_w}\right)+
\frac{1-S_w}{\mu_n}\ \left(\frac{d}{dS_w}\right)
\left(\frac{k_{r,n}}{1-S_w}\right)=0\;,
\end{equation}
and
\begin{equation}
\label{eqn33}
\left(\frac{d}{dS_w}\right) \left(\frac{k_{r,w}}{S_w}\ P_c\right)=0\;.
\end{equation}
This latter equation may in turn be integrated to give
\begin{equation}
\label{eqn34}
P_c=P_0\ \frac{S_w}{k_{r,w}}\;,
\end{equation}
where $P_0$ is a reference pressure.

We see here that the fundamental assumptions of relative permeability 
concept are challenged: equation (\ref{eqn32}) {\it contains the 
viscosities\/} $\mu_w$ and $\mu_n$.  Hence, at least one of the relative
permeabilities $k_{r,w}$ and $k_{r,n}$ {\it must\/} depend of the viscosity
ratio $\mu_n/\mu_w$.

We also see from equation (\ref{eqn34}) that the capillary pressure $P_c$ 
{\it cannot become negative.\/} However, under mixed wetting conditions, 
it does \cite{abcehfghhmz86}. 

There are equations relating the capillary pressure to the 
relative permeability in the literature such as the Brooks-Corey relation 
\cite{bc64} or the van Genuchten equation \cite{g80}.  These equations all 
build on the work of Purcell \cite{p49} who generalized to pore networks 
the relation between  the capillary pressure and the permeability of a single 
capillary tube via the Young formula for interfacial tension.  Equation 
(\ref{eqn34}) is different from these earlier equations in that it does not 
rely on assumptions concerning the physics of the problem apart from the
balance of dissipation in and work on the fluids.  We also note that popular 
parametrizations of the relative permeabilities such as that of Corey 
\cite{c54} and that of Lomeland et al.\ \cite{leh05} (``LET"-type) do not 
obey equation (\ref{eqn32}).  Hence, they are thermodynamically inconsistent,
as are the basic assumption that the relative permeabilities $r_{r,w}$ 
and $k_{r,n}$ are functions of the saturation $S_w$ alone.

\section{Discussion and conclusion}
\label{conclusion}

We have in this paper constructed a theory based on non-equilibrium
thermodynamics that reduces the immiscible {\it two-phase flow\/} in porous 
media to a {\it one-phase\/} flow problem.  Let us now consider a 
three-dimensional isotropic porous medium.  Let $\vec x$ be a point somewhere
in this porous medium. The theory that we have developed can then be
summarized by the following set of equations, 
\begin{equation}
\label{conc-1}
\frac{\partial S_w}{\partial t}+\vec\nabla\cdot (S_w\vec v_w)=0\;,
\end{equation}
\begin{equation}
\label{conc-5}
\vec v = S_w \vec v_w+(1-S_w)\vec v_n\;,
\end{equation}
and
\begin{equation}
\label{conc-6}
S_w\frac{d\vec v_w}{dS_w}+ (1-S_w)\frac{d\vec v_n}{dS_w}=0\;.
\end{equation}
Here (\ref{conc-1}) is the conservation law for the wetting saturation. This
expression becomes the Buckley-Leverett equation \cite{bl42} if
we set $S_w\vec v_w=F_w\vec v$, see equation (\ref{eqn21}) and
take the incompressibility of the fluids into account. The
non-wetting saturation $S_n$ has been eliminated by using the 
incompressibility of the two fluids.  This implies that 
$\vec\nabla\cdot\vec v=0$ is
build into the equation set.  Equation (\ref{conc-6})
is the three-dimensional version of equation (\ref{eqn20}). The 
three-dimensional version of equation (\ref{eqn18}) is
$d\vec v/dS_w=\vec v_w-\vec v_n$.  This equation follows by taking the
derivative of (\ref{conc-5}) with respect to $S_w$ and using (\ref{conc-6}).
Hence, this equation is also contained in (\ref{conc-1}) to (\ref{conc-6}).  It
may replace either of the equations (\ref{conc-5}) or (\ref{conc-6}).
These equations are all conservation
laws, the last two express power input equals dissipation.
Thus, they transcend the details of the porous medium.  Equations
(\ref{conc-1}) to (\ref{conc-6}) are 7 
equations.  There are ten variables $S_w$, $\vec v$, $\vec v_w$ and $\vec v_n$.
The three equations that close the system of equations are
the {\it constitutive equations\/} 
\begin{equation}
\label{conc-7}
\vec v=\vec v[\vec x,S_w(\vec x),\vec\nabla P(\vec x)]\;,
\end{equation}
containing the detailed physics of the system.

We see that the constitutive equation does not contain the seepage
velocities of the immiscible fluids, $\vec v_w$  and $\vec v_n$, explicitly.
Only the saturation $S_w$ enters.  Hence, the constitutive equation
(\ref{conc-7}) can be interpreted as that of a single fluid depending
on one extra variable, $S_w(\vec x)$.  Hence, the equation set (\ref{conc-1})
--- (\ref{conc-7}) reduces the immiscible two-phase flow problem in porous
media to a one-phase flow problem involving a complex fluid.
This viewpoint permeates recent 
work on the effective permeability of immicible two-phase systems
where it is suggested that two fluids behave as if they were a single 
Bingham plastic \cite{tkrlmtf09,tlkrfm09,rcs11,sh12,sbdkpbtsch16}.

We may clarify this point even further by eliminating the two fluid
velocities $\vec v_w$ and $\vec v_n$ in equations (\ref{conc-1}) to 
(\ref{conc-6}).  The equation set then reduces to a single equation
\begin{equation}
\label{conc-8}
\frac{\partial S_w}{\partial t}+\vec v\cdot\vec\nabla S_w=S_w
\left(\frac{d\vec v}{dS_w}\right)\cdot\vec\nabla S_w\;.
\end{equation}
Together with the constitutive equation (\ref{conc-7}), we now have a
closed set describing effectively a single fluid with a velocity
field $\vec v$ which transports an active scalar $S_w$. 

The velocities of the two immiscible fluids may then be found by using 
the equations
\begin{equation}
\label{conc-9}
\vec v_w=\vec v +(1-S_w)\frac{d\vec v}{dS_w}\;,
\end{equation}
and
\begin{equation}
\label{conc-10}
\vec v_n=\vec v -S_w\frac{d\vec v}{dS_w}\;,
\end{equation}
which may be derived from equation (\ref{conc-5}) and (\ref{conc-6}).

In this manuscript we have considered a single driving force that induces
the flow: the pressure gradient.  Other forces such as buoyancy, temperature
gradients and chemical driving forces can be incorporated in the 
non-equil-ibrium thermodynamics formalism used in section \ref{diss}.  The
same is true for the introduction of more immiscible fluids than two.
Hence, equations (\ref{conc-1}) to (\ref{conc-6}) may be generalized
to include these additional complications.

\begin{acknowledgements}
The authors thank Eirik Grude Flekk{\o}y, Knut J{\o}rgen M{\aa}l{\o}y, 
Tho\-mas Ramstad, Per Arne Slotte and Marios Valavanides for interesting 
discussions on this topic.  AH, SK and IS thank VISTA, a collaboration between 
Statoil and the Norwegian Acad\-emy of Sciences, for financial support.  
SS thanks the Norwegian Research Council, NFR and the Beijing 
Computational Science Research Center CSRC for financial support. 
\end{acknowledgements}

\bibliographystyle{spmpsci}

\begin{thebibliography}{}

\bibitem{b72} J.\ Bear, Dynamics of fluids in porous media, 
Dover, Mineola, 1988.

\bibitem{wb36} R.\ D. Wyckoff and H.\ G.\ Botset,  The flow of gas-liquid
mixtures through unconsolidated sands, J.\ Appl.\ Phys.\ {\bf 7}, 325 (1936).

\bibitem{r31} L.\ A.\ Richards, Capillary conduction of liquids through
porous mediums, J.\ Appl.\ Phys.\ {\bf 1}, 318 (1931).

\bibitem{l40} M.\ C.\ Leverett, Capillary behavior in porous sands, Trans.\
AIMME, {\bf 12}, 152 (1940).

\bibitem{lsd81} R.\ G.\ Larson, L.\ E.\ Scriven and H.\ T.\ Davis, Percolation
theory of two phase flow in porous media, Chem.\ Eng.\ Sci.\ {\bf 36}, 57
(1981).

\bibitem{hg90} S.\ M.\ Hassanizadeh and W.\ G.\ Grey, Mechanics and 
thermodynamics of multiphase flow in porous media including interphase
boundaries, Adv.\ Wat.\ Res.\ {\bf 13}, 169 (1990).

\bibitem{hg93a} S.\ M.\ Hassanizadeh and W.\ G.\ Grey, Towards an improved
description of the physics of two-phase flow, Adv.\ Wat.\ Res.\ {\bf 16}, 53 
(1993).

\bibitem{hg93b} S.\ M.\ Hassanizadeh and W.\ G.\ Grey, Thermodynamic basis
of capillary pressure in porous media, Wat.\ Res.\ Res.\
{\bf 29}, 3389 (1993).

\bibitem{h98} R.\ Hilfer, Macroscopic equations of motion for two-phase 
flow in porous media, Phys.\ Rev.\ E, {\bf 58}, 2090 (1998).

\bibitem{hb00} R.\ Hilfer and H.\ Besserer, Macroscopic two-phase flow in
porous media, Physica B, {\bf 279}, 125 (2000).

\bibitem{h06a} R.\ Hilfer, Capillary pressure, hysteresis and residual
saturation in porous media, Physica A, {\bf 359}, 119 (2006).

\bibitem{h06b} R.\ Hilfer, Macroscopic capillarity and hysteresis for flow
in porous media, Phys.\ Rev.\ E, {\bf 73}, 016307 (2006).

\bibitem{h06c} R.\ Hilfer, Macroscopic capillarity without a constitutive
capillary pressure function, Physica A, {\bf 371}, 209 (2006).

\bibitem{hd10} R.\ Hilfer and F.\ D{\"o}ster, Percolation as a basic
concept for capillarity, Transp.\ Por.\ Med.\ {\bf 82},
507 (2010).

\bibitem{nbh11} J.\ Niessner, S.\ Berg and S.\ M.\ Hassanizadeh, Comparison
of two-phase Darcy's law with a thermodynamically consistent approach,
Transp.\ Por.\ Med.\ {\bf 88}, 133 (2011).

\bibitem{dhh12} F.\ D{\"o}ster, O.\ H{\"o}nig and R.\ Hilfer, Horizontal
flow and capillarity-driven redistribution in porous media, Phys.\ Rev.\
E, {\bf 86}, 016317 (2012).

\bibitem{habgo15} R.\ Hilfer, R.\ T.\ Armstrong, S.\ Berg, A.\ Georgiadis 
and H.\ Ott, Phys.\ Capillary saturation and desaturation, Rev. E, {\bf 92}, 
063023 (2015).

\bibitem{h15} S.\ M.\ Hassanizadeh, Advanced theories for two-phase flow
in porous media, in Handbook of Porous Media, 3rd edition, edited by K.\ 
Vafai, CRC Press, Boca Raton, 2015.

\bibitem{gsd16} B.\ Ghanbarian, M.\ Sahimi and H.\ Daigle, Modeling relative 
permeability of water in soil: Application of effective-medium 
approximation and percolation theory, Water Res.\ Res.\ {\bf 52}, 5025 (2016).

\bibitem{bl42} S. E. Buckley and M. C. Leverett, Mechanism of fluid
displacements in sands, Trans. AIME, {\bf 146}, 107 (1942).

\bibitem{kp98} D.\ Kondepudi and I.\ Prigogine, Modern thermodynamics, Wiley, 
Chichester, 1998.

\bibitem{kb08} S. Kjelstrup and D. Bedeaux, Non-equilibrium thermodynamics
for heterogeneous systems, World Scientific, Singapore, 2008.

\bibitem{kbjg10} S.\ Kjelstrup, D.\ Bedeaux, E.\ Johannesen and J.\ Gross,
Non-equilibrium thermodynamics for engineers, World Scientific,
Singapore, 2010.

\bibitem{amhb98} E.\ Aker, K.\ J.\ M{\aa}l{\o}y, A.\ Hansen and G.\ G.\ 
Batrouni, A two-dimensional network simulator for two-phase flow in porous 
media, Transp.\ Porous Media, {\bf 32}, 163 (1998). 

\bibitem{tkrlmtf09} K.\ T.\ Tallakstad, H.\ A.\ Knudsen, T.\ Ramstad, G.\ 
L{\o}voll, K.\ J.\ M{\aa}l{\o}y, R.\ Toussaint and E.\ G.\ Flekk{\o}y, 
Steady-state two-phase flow in porous media: statistics and transport 
properties, Phys.\ Rev.\ Lett.\ {\bf 102}, 074502 (2009).

\bibitem{tlkrfm09} K.\ T.\ Tallakstad, G.\ L{\o}voll, H.\ A.\ Knudsen, T.\ 
Ramstad, E.\ G.\ Flekk{\o}y and K.\ J.\ M{\aa}l{\o}y, Steady-state 
simultaneous two-phase flow in porous media: an experimental study, Phys.\ 
Rev.\ E {\bf 80}, 036308 (2009).

\bibitem{rcs11} E.\ M.\ Rassi, S.\ L.\ Codd and J.\ D.\ Seymour, Nuclear 
magnetic resonance characterization of the stationary dynamics of partially 
saturated media during steady-state infiltration flow, N.\ J.\ Phys.\ 
{\bf 13}, 015007 (2011).

\bibitem{sh12} S.\ Sinha and A.\ Hansen, Effective rheology of immiscible 
two-phase flow in porous media, Europhys.\ Lett.\ {\bf 99}, 44004 (2012).

\bibitem{sbdkpbtsch16} S. Sinha, A. T. Bender, M. Danczyk, K. Keepseagle,
C. A. Prather, J. M. Bray, L. W. Thrane, J. D. Seymor, S. L. Codd and
A. Hansen, Effective rheology of two-phase flow in three-dimensional porous 
media: experiment and simulation, submitted to Trans. Por. Med. (2016).

\bibitem{yts13} A.\ G.\ Yiotis, L.\ Talon and D.\ Salin, Blob population 
dynamics during immiscible two-phase flow in reconstructed porous media, 
Phys.\ Rev. E, {\bf 87}, 033001 (2013).

\bibitem{bt13} R.\ G.\ Bentsen and J.\ Trivedi, On the construction of an 
experimentally based set of equations to describe cocurrent or 
countercurrent, two-phase flow of immiscible fluids
through porous media, Transp.\ in Por.\ Media, {\bf 99}, 251 (2013).

\bibitem{mfmst15} M.\ Moura, E.\ -A.\ Florentino, K.\ J.\ M{\aa}l{\o}y, G.\ 
Sch{\"a}fer and R.\ Toussaint, Impact of sample geometry on the measurement 
of pressure-saturation curves: experiments and simulations, Water Res.\ Res.\ 
{\bf 51}, 8900 (2015).

\bibitem{abcehfghhmz86} W.\ Abdallah, J.\ S.\ Buckley, A.\ Carnegie,
J.\ Edwards, E.\ Fordham, A.\ Graue, T.\ Habashy, H.\ Husain, B.\ Montaron
and M.\ Ziauddin, Fundamentals of wettability, Techology, {\bf 38}, 1125
(1986).

\bibitem{bc64} R.\ H.\ Brooks and A.\ T.\ Corey, Hydraulic properties of 
porous media, Colorado State University hydrology paper {\bf 3}, Colorado 
State University, 1964.

\bibitem{g80} M.\ Th.\ van Genuchten, A closed-form equation for predicting 
the hydraulic conductivity of unsaturated soils, Soil Sci.\ Soc.\ Am.\ J.\ 
{\bf 44}, 892 (1980).

\bibitem{p49} W.\ R.\ Purcell, Capillary pressure --- their measurement using 
mercury and the calculation of permeability therefrom, Journal Petr.\ Tech.\ 
{\bf 1}, 39 (1949). 

\bibitem{c54} A.\ T.\ Corey, The interrelation between gas and oil relative 
permeabilities, Prod.\ Monthly, {\bf 19}, 38 (1954).

\bibitem{leh05} F.\ Lomeland, E.\ Ebeltoft and T.\ W.\ Hammervold, A new 
versatile relative permeability correlation, Rev.\ Proc. of the 2005 Int.\ 
Symp.\ of the SCA, Abu Dhabi, Oct.\ 31 -- Nov.\ 2, 2005.

\end{thebibliography}

\end{document}